\documentclass[twocolumn]{revtex4-1}

\usepackage[utf8]{inputenc}

\usepackage{graphicx}
\usepackage{subfigure}
\usepackage{float}

\usepackage{amsmath}
\usepackage{amssymb}
\usepackage{amsfonts}
\usepackage{amsthm}
\usepackage{dsfont}
\usepackage{color}

\newcommand{\rs}{\rm \scriptscriptstyle}

\begin{document}

\title{Fractional Excitations in Cold Atomic Gases}

\author{J. Honer${}^1$}
\author{J. C. Halimeh${}^2$}
\author{Ian McCulloch${}^3$}
\author{U. Schollw\"ock${}^2$} 
\author{H. P. B\"uchler${}^1$}

\affiliation{$^{1}$Institute for Theoretical Physics III, University of Stuttgart, Germany}
\affiliation{${}^2$Department of Physics and Arnold Sommerfeld Center for Theoretical Physics, Ludwig-Maximilians-Universit\"at M\"unchen, Germany}
\affiliation{${}^3$Centre for Engineered Quantum Systems, School of Mathematics and Physics, University of Queensland, Brisbane 4072, Australia}

\date{\today}

\begin{abstract}

We study the behavior of excitations in the tilted one-dimensional Bose-Hubbard model.
In the phase with broken symmetry, fundamental excitations are domain-walls which show fractional statistics.
Using perturbation theory, we derive an analytic model for the time evolution of these fractional excitations, and demonstrate the existence of a repulsively bound state above a critical center of mass momentum. 
The validity of the perturbative analysis is confirmed by the use of t-DMRG simulations. 
These findings open the path for experimental detection of fractional particles in cold atomic gases.

\end{abstract}

\maketitle

Excitations carrying fractional quantum numbers are one of the most intriguing features of strongly interacting many-body systems. Arguably the most celebrated of those is the charge $e/3$ Laughlin quasiparticle responsible for the fractional quantum hall effect. Much effort is devoted to finding novel phases with even more exotic excitations such as non-abelian anyons and fractional statistics in three-dimensional systems. In the quest for the experimental realization of such strongly correlated phases, cold atomic gases with their  clean and controllable environment are a promising candidate and testing ground  \cite{doi:10.1080/00018730701223200}. A first step in this direction marks the recently observed phase transition in a one-dimensional cold atomic system \cite{nature09994}, where  the fundamental  excitations in the broken-symmetry phase  are domain walls carrying an effective fractional charge. In this letter, we investigate the creation and detection of these fractional excitations in the  one-dimensional tilted Bose-Hubbard model.

\begin{figure}[t]
\includegraphics[width= .8\columnwidth]{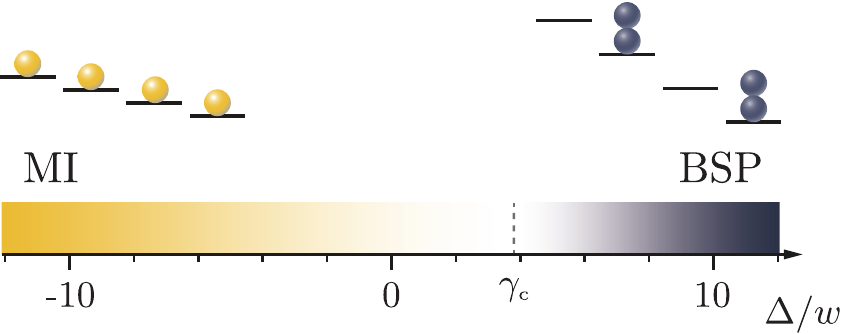}
\caption{
Phase diagram for the tilted Bose-Hubbard model in the regime $\Delta, w \ll U, E >0$ and an averaged density of one particle per lattice site: 
Mott insulator (MI) ground state for $\Delta/w< \gamma_c \approx 1.85$ \cite{PhysRevB.66.075128}, and the broken symmetry phase (BSP) for $\Delta/w >\gamma_c$.
}
\label{fig1}
\end{figure}

A variety of theoretical proposals in cold atomic gases focus on the realization of  strongly correlated phases with  fractional excitations, such as  models supporting spin  liquid phases \cite{PhysRevLett.95.040402}, as well as  Kitaev's toric code with abelian anyonic excitations \cite{nphys287, PhysRevLett.91.090402, PhysRevLett.98.150404}, and systems  in large effective magnetic fields \cite{RevModPhys.83.1523, 1367-2630-5-1-356} exhibiting fractional quantum hall states \cite{srep00043, PhysRevLett.96.180407, ncomms1380, 1367-2630-14-5-055003, PhysRevA.70.053612, PhysRevLett.94.086803, PhysRevLett.94.070404}.  These proposals are based on standard experimental tools nowadays available in the context of quantum simulation, see Ref.~\cite{nphys2259} for a review.  Especially, the latest development of experiments \cite{nature08482, nature09378} with single-site readout and addressability in optical lattices  has opened the path for the observation of a novel quantum phase transition in tilted optical lattices \cite{nature09994, PhysRevB.66.075128, PhysRevLett.109.015701,PhysRevB.83.205135}. The transition in this one-dimensional system takes place from a phase with one atom on each lattice site, 
to a broken-symmetry phase (BSP) with an alternating mean occupation, see Fig.~\ref{fig1}. 

Here, we study the behavior of excitations in the tilted one-dimensional Bose-Hubbard model (TBH) in the phase with broken symmetry, with the latter being two-fold degenerate.  Fundamental excitations are domain walls between the degenerate broken-symmetry phases and exhibit an effective fractional charge.
In turn, the simplest excitation corresponds to moving a single atom from one site to its neighboring site.  Such an excitation corresponds to the creation of two closely bound domain walls. 
The important question is then, whether these experimentally accessible excitations will decay into the fundamental  domain-wall excitations and how to detect these fractional excitations. We derive an analytic expression for the time evolution of those excitations, and show the existence of a repulsively  bound state of fractional excitations above a critical center of mass momentum $Q_{\rs c} = 2\pi/3 $. Further we provide  experimental signatures for measuring  fractional excitations in a setup of cold atoms and give direct numerical calculations for the time evolution for finite size sample.

\begin{figure}[h]
\includegraphics[width= .8\columnwidth]{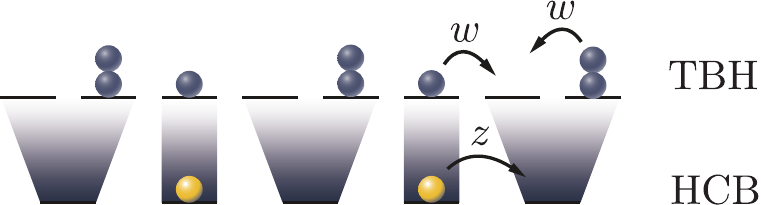}
\caption{
Mapping from the tilted Bose-Hubbard model (TBH) onto the effective hard-core boson model (HCB): two neighboring lattice sites occupied by a $0,2$-configuration map onto a single unoccupied site,
whereas a singly occupied site  maps onto an occupied  site. 
A second-order transition with $2 w^2/\Delta$ allows for an effective hopping of the fractional excitations.
}
\label{fig2}
\end{figure}

We start with the one-dimensional tilted Bose-Hubbard model as realized in the 
experimental setup \cite{nature08482}, which takes the form
\begin{align}
\begin{aligned}
H_{\rs TBH} =& -w \sum_i \left( a_i^\dagger a_{i+1} + a_i a^\dagger_{i+1}\right) \\&
+ \frac{U}{2} \sum_i n_i(n_i-1) - E \sum_i i n_i
\label{TBH}
\end{aligned}
\end{align}
with the on-site interaction $U$, the hopping rate $w$, and the  lattice tilt $E$ per site.  In addition, $a_i^\dagger$ ($a_i$) is the creation (annihilation) operator for a particle on site $i$, and $n_i = a^\dagger_i a_i$  is the number operator on site $i$. We focus on the regime with an averaged density of one particle per lattice site, and assume a  positive tilt $E>0$, i.e. the lattice is tilted to the right. The system is in a meta-stable state, where the condition that the on-site interaction $U$ and the lattice tilt $E$ are much larger than the tunneling energy $w$,  prevents the relaxation into a state with more than two-particles accumulating in a single site. Then, the relevant energies are the energy difference $\Delta = E-U$ and the hopping rate $w$.

For large negative $\Delta$, the ground state is a Mott insulator (MI) with one particle per lattice site. Increasing $|\Delta|\sim w$, particles can 
tunnel to their right neighboring site, as long as the particle on that site has not tunneled. Eventually, the system undergoes
a phase transition into a ground state with broken translational symmetry with two atoms on each second lattice site, see Fig.~\ref{fig1}.

The condition $E,U \gg w, \Delta$ suppresses processes with  particles hopping to the left as well as the accumulation of more than two particles in a single lattice site. These constraints are most conveniently incorporated by a mapping of the Bose-Hubbard model to a spin model:
The spin degree of freedom resides between two lattice sites $i$ and $i+1$ and the spin-up state corresponds to a particle at lattice site $i$, while the spin-down state accounts for the situation where the particle has tunneled to the site $i+1$.
Then, the Bose-Hubbard model maps to   a one-dimensional Ising model
\begin{align}
\begin{aligned}
H_{\rs Ising} =& - 2 \sqrt{2} w  \sum_i S_x^{i}  + \Delta \sum_i S_z^{i} \\
&+ J \sum_i \left(S_z^{i} - \frac{1}{2}\right)\left(S_z^{i+1} - \frac{1}{2}\right),
\end{aligned}
\end{align}
with $S_z^i$ ($S_x^i$) is the spin operator along the $z$ ($x$) axis. In addition, the last term accounts for the constraint that a particle can only tunnel from site $i$  to $i+1$ if there is already a particle at site $i+1$ and formally requires taking the limit $J\rightarrow \infty$.  Under this mapping, the Mott-insulating state corresponds to a paramagnetic phase with all spin-up, while the ordered broken-symmetry phase corresponds to an Ising antiferromagnetic ground state.

In the following, we focus on the quantum phase for $\Delta \gg w$ with two particles on every second lattice site.  The experimentally accessible excitations are achieved by moving a particle from one lattice site to its neighboring site to the left,  which corresponds to a single spin-flip.
The energy of this excitation is given by $\Delta$. This single-particle excitation can now be decomposed into two fractional excitations via a second order-process as indicated in Fig.~\ref{fig2}. 
In the classical regime with $w=0$, these delocalized fractional excitations have the same energy $\Delta$ as the single-spin excitations, while
adding quantum fluctuations within second-order perturbation theory yields a nearest-neighbor interaction  $V= 2 w^2/\Delta$, as well as
an effective hopping  for the fractional excitations with  amplitude $z = 2 w^2/\Delta$. 
Note that fractional excitations are restricted to even or odd lattice sites, depending on the local order, and hence hopping takes place from site $i$ to site $i\pm2$. This restriction can be uniquely fulfilled by introducing a new lattice:
each singly occupied site and the combination of a doubly occupied site next to an empty site count as a new lattice site (see Fig.~\ref{fig2}). Consequently, the number of lattice sites between two excitations is halved, and the dynamics of fractional excitations is governed by the hard-core boson model (HCB) with nearest-neighbor hopping and interaction $z$ and $V$, respectively, described by the Hamiltonian 
\begin{align}
H_{\rs HCB} &= - z \sum_j \left( b_j^\dagger b_{j+1} + b_{j} b_{j+1}^{\dagger} \right)+ V \sum_j n_j n_{j+1}.
\label{HCBH}
\end{align}
Here  $b_j^\dagger$ ($b_j$) is the creation (annihilation) operator for a fractional excitation on site $j$, and $n_j = b^\dagger_j b_j$,  is the number operator with $\langle n_j \rangle \in [0,1]$.

A similar Hamiltonian has previously been studied in the context of repulsively bound pairs \cite{nature04918, PhysRevA.81.045601}. 
Here, a bound state of domain walls means that  the experimental excitation cannot decay into its  fractional parts due to energy conservation,
while the absence of a bound state corresponds to a situation where this experimental excitation decays into delocalized fractional excitations and  allows for their experimental observation. As in the present situation, $z=V$ is the only energy scale of the model, we obtain a universal bound-state structure for the fractional excitations, which we  analyze in the following.

We write the wave-function for two fractional excitations in the effective model as $\psi(i,j)$, where $i$ and $j$ denote the positions of the fractional excitations in the effective lattice.
The discrete translational invariance of the system provides conservation of the center of mass quasi-momentum, which allows for an expansion into eigenfunctions of $H_{\rs HCB}$ for fixed center of mass quasi-momentum $Q$. Therefore, the two-particle wave-function can be factorized as $\psi(X,x) = e^{- i Q X} \psi (x)$ with the center of mass $X = (i+j)/2$ and relative coordinates $x=i-j$.
We find two different regimes for the two-particle states:
In the first regime for $|Q| \le Q_{\rs c} = 2 \pi /3$,
the two-particle eigenfunctions are given by scattering states $\psi_{q}$ alone with energy  $E_{q} = -  2 z_Q \cos q$, where $q$ is the relative momentum and 
$z_Q = 2 z  \cos Q/2$ denotes the hopping rate in the center of mass frame. 
Its wave-function reduces to plane waves, $\psi_{q}(x) =  (1-\delta_{x,0}) \cos(q |x| + \phi_{Q,q})$, with a scattering phase shift 
 \begin{align}
\phi_{Q,q} &= \arctan \frac{\cos q + 2 \cos Q/2}{\sin q} .
\notag
\end{align}
The Kronecker-delta factor $1-\delta_{x,0}$ accounts for the hard-core constraint and enforces $\psi_{q} = 0$ at $x=0$. 
In the second regime for $|Q| > Q_{\rs c }$, an additional bound state $\psi_{\rs B}$ emerges. The repulsive interaction yields an energy  $E_{\rs B} = z (1-4 \cos^2 Q/2) $ lying above the scattering continuum.
Its two-particle wave-function shows an exponential decay with relative distance $x$, and can be written as
\begin{align}
\psi_{\rs B} (x) &=   (1-\delta_{x,0})    \left[\frac{1 - 4 \cos^2  Q/2}{4 \cos^2  Q/2}\right]^{\frac{1}{2}}
\left( -2 \cos \frac{Q}{2}\right)^{|x|}  \!\!\!\!\!.
\notag
\end{align}
Note that the alternating amplitude of the wave-function is a typical feature of a repulsively bound state. 
The general wave-function of two fractional excitations with fixed center of mass momentum $Q$ can therefore be decomposed as
\begin{align}
\psi (x)=   C_{\rs B}\psi_{\rs B} e^{- i E_{\rs B} t/\hbar} + \int  \frac{dq}{2 \pi}  C_{q}\psi_{q} e^{- i E_{q} t/\hbar}
\label{decomp}
\end{align}
with 
$C_{\rs B}$ and $C_{q}$
denoting the overlap 
of the initial wave-function 
with the bound and scattering eigenfunctions of $H_{\rs HCB}$.  
The experimentally accessible initial states are achieved by moving one particle to the left, which in the effective model
corresponds to two adjacent occupied sites $\psi_{\rs SPE} (x) = \delta_{x,1}$. 
Then the overlap is given by $C_{\rs B}=  \theta(|Q| - Q_{\rs c}) \sqrt{1-4 \cos^2 Q/2} $ and $C_{q} = \sqrt{2} \cos(q + \phi_{Q,q})$, with  $\theta(x)$ being the Heaviside step function.  
The integral over the relative quasi-momentum $q$ in eq.~(\ref{decomp}) can be carried out analytically, giving rise to a formal solution in terms of an infinite sum of Bessel-functions of the first kind
\begin{align}
\int  \frac{dq}{2 \pi}  C_{q}\psi_{q}(x) e^{- i E_{q} t/\hbar} = \sum_n c_n(x) e^{i \frac{\pi}{2}n} J_n (2 t z_Q/\hbar) 
\label{Bessel}
\end{align}
with coefficients $c_n(x)$ defined via a discrete Fourier transform $C_{q} \psi_{q}(x)  = \sum_n c_n(x) e^{i q n}$. 

\begin{figure}[htbp]
\includegraphics[width= .8\columnwidth]{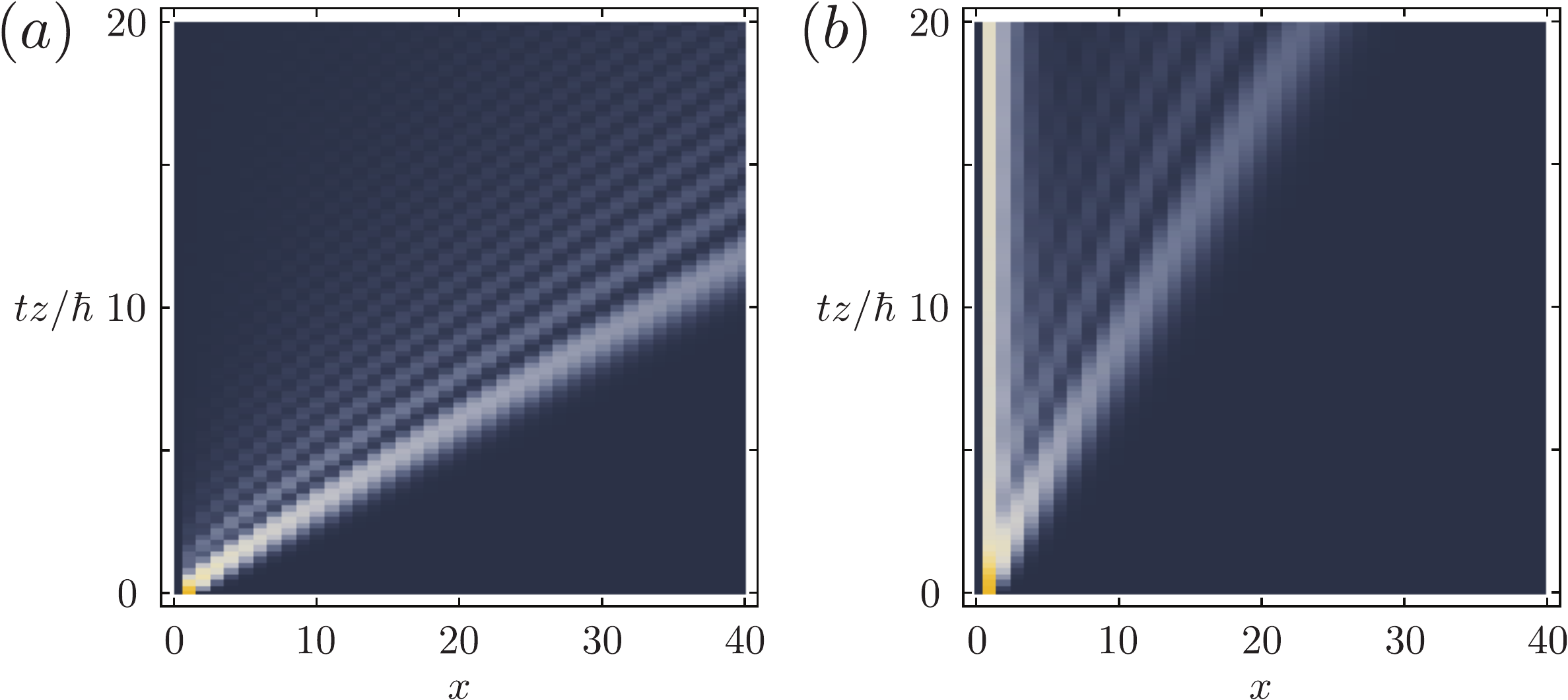}
\caption{
Time evolution of the relative wave-function $\psi (x)$:
(a) For $|Q| \le Q_{\rs c}$, the wave-function is strongly localized at around a relative distance $x \sim t z_Q/\hbar$. 
(b) For $|Q| > Q_{\rs c}$, the 
bound state adds an additional exponentially decaying contribution, with its maximum at $x=1$.
}
\label{fig4}
\end{figure}

\begin{figure}[t]
\includegraphics[width= .8\columnwidth]{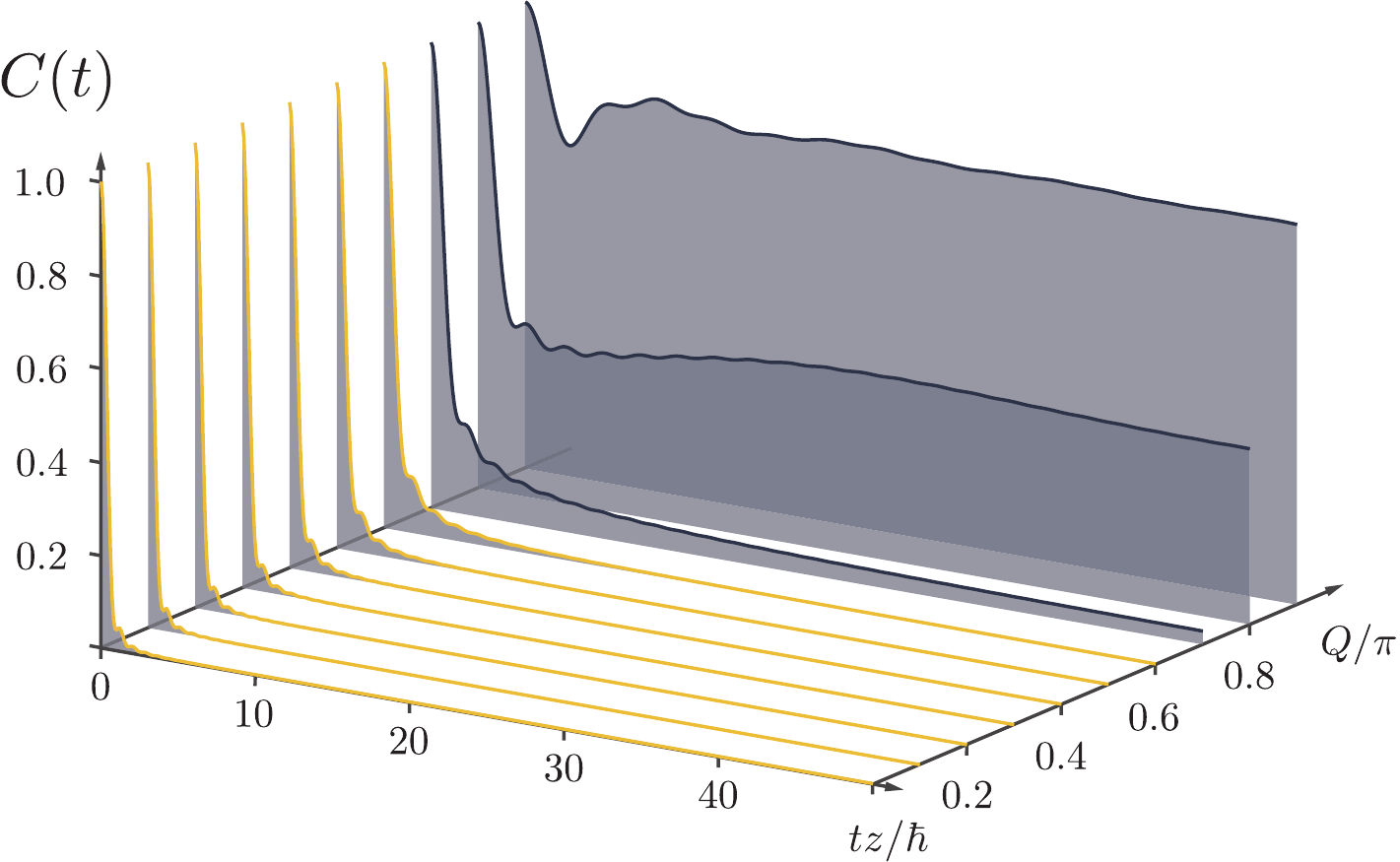}
\caption{Time evolution of the correlation function $C(t)$ for different center of mass quasi-momenta $Q$:
Below the critical momentum $|Q|<Q_{\rs c}$, the correlation function decays to zero, while for $|Q|>Q_{\rs c}$, the overlap with the emerging repulsively bound state gives a finite probability for fractional excitations to stay at finite relative distance $x$, resulting in a finite value of the correlation function for $t z_Q/\hbar \gg 1$. }
\label{fig5}
\end{figure}

The time evolution of the wave-function is shown in Fig.~\ref{fig4}. The superposition of the scattering states leads to a ballistic expansion
of the fractional excitations with a velocity determined by the hopping energy $z_{Q}$, i.e., the two-particle wave-function $\psi_{\rs SPE}$ is strongly 
localized around a linearly growing relative distance  $x \sim t z_Q /\hbar$, with some additional interference fringes appearing at smaller relative 
distances $x$, but propagating at the same velocity. However, the finite overlap with the bound state $\psi_{\rs B}$ for $Q > Q_{\rs c}$ creates an 
additional stationary peak at $x \sim 1$ (see Fig.~\ref{fig4}(b)).  With the scattering states moving away from each other, 
 measurement of the wave-function amplitude at $x=1$  at times $t \gg 1/z_Q$  allows one to single out the bound-state contribution. 
Formally, this can be cast in terms of a correlation function 
$C(t) = |\psi_{\rs SPE} (X,x=1)|^2$ on the effective lattice. In the microscopic lattice, $C(t)$ corresponds to $\langle P_i P_{i+1}\rangle$, with $P_{i}=n_{i}(n_{i}-2) $ being the projection operator on singly occupied sites.
The time evolution of $C(t)$ for different center of mass quasi-momenta is shown in Fig.~\ref{fig5}. 
For $Q<Q_{c}$, the correlation function decays to zero with a characteristic behavior $ \sim (t  z_Q/\hbar)^{-3}$, and exhibits characteristic oscillations 
due to interference between the different scattering states.  
In addition, the decay exhibits an intermediate regime with a characteristic behavior 
 $\sim (t z_Q/\hbar)^{-1}$. The time scale for the crossover between the long-time behavior and this intermediate regime diverges approaching
 the critical value  $Q_{c}$.  Consequently, the decay at $Q=Q_{c}$ is given by a critical behavior   $\sim (t z_Q/\hbar)^{-1}$ for the correlation function, which
 follows from the analytical expression for the wave-function
\begin{align}
\begin{aligned}
\psi_{\rs SPE} (x)=& e^{-i \frac{\pi}{2} |x|} \left[   J_{|x|} (2 z t ) + i J_{|x|-1} (2 z t ) \right] .
\notag
\end{aligned}
\end{align}
Finally, the presence of a bound-state is characterized by a saturation of the correlation function 
\begin{align}
C(t) \rightarrow \theta(|Q| - Q_{\rs c}) \left[1-4 \cos^2 Q/2\right]^2
\notag
\end{align}
for   $t z_Q/\hbar \gg 1$.  The bound-state contribution grows steadily towards the edge of the Brillouin zone, and
eventually,  $\psi_{\rs SPE}$ becomes an eigenstate of the Hamiltonian at $|Q|=\pi$ with a constant correlation function $C(t) = 1$. 
In addition, the interference between the bound state and the scattering states gives rise to a
characteristic beating of the correlation function with the frequencies $\omega^\pm = E_{\rs B} \pm E_{\rs SC}(q=0) = z (1 - 4\cos^2 Q/2 \pm 2 \cos Q/2)$.
This gives experimental access to the energy gap between bound and scattering states, and allows for the measurement of the bound-state energy.

\begin{figure}[hbt]
\includegraphics[width= .8\columnwidth]{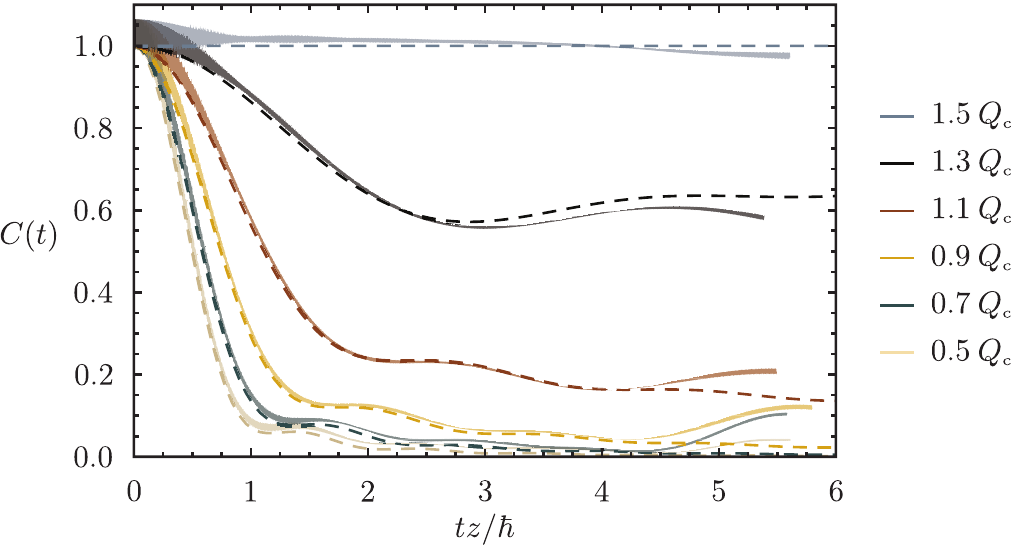}
\caption{t-DMRG results (solid lines) in a finite lattice with $L=30$ microscopic sites (i.e. $l=16$ sites in the effective lattice) in comparison to the analytic results (dashed lines) in an infinite lattice for various center of mass momenta $Q$. Finite size effects start come into play at times $t \sim 4 \hbar/z$.
}
\label{fig6}
\end{figure}

To confirm our perturbative results and to provide further insight into an experimental realization, we provide time-adaptive density matrix renormalization group (t-DMRG) simulations \cite{1742-5468-2004-04-P04005, PhysRevLett.93.076401}
of a single-particle excitation in the original tilted Bose-Hubbard model in eq.~(\ref{TBH}). 
Here, we use a realistically large lattice of $L=30$ sites, i.e. $ l = 16$ sites in the effective lattice.
Time-evolution calculations are performed using second- and fourth-order Trotter decompositions. 
A comparison between the analytic correlation function 
and the  t-DMRG result is shown in Fig.~\ref{fig6}.

On the one hand, t-DMRG results show good agreement with the perturbative model. We find that the timescale of the correlation-function decay and the saturation values 
agree well with the analytic results. 

On the other hand, we observe deviations due to finite system size and finite values $U$, $E$, and $w$, which we  discuss in the following.
First, 
the simulations are performed in a finite size system. This leads to revivals of the correlation function $C(t)$ due to scattering of fractional excitations at the system boundaries. In a microscopic lattice of $L=30$ sites, we find deviations due to finite size at times $t \sim 4 \hbar/z$.
A hard-core  model neglecting  nearest-neighbor interaction, which allows for an analytic solution even in a finite system, agrees well with the t-DMRG result and gives an estimate for revival times. From this, we can derive a lower boundary of $L\gtrsim26$  to observe a signature for a bound state in the correlation function.
Second, the simulated correlation function shows an additional high-frequency oscillation. This can be explained as follows: 
Since the initial state is not an energy eigenstate of the Hamiltonian for a finite ratio $w/\Delta$, this leads to interference between different energy contributions, and ultimately in oscillations on the order of the excitation spectrum, i.e. $\Delta$. We find that for a ratio of $\Delta/w \gtrsim 30$  the suppression of these oscillations is strong enough to see a clear correlation-function signal (see Fig.~\ref{fig6}). 

Finally, the validity of the correlation function is based on the stability of the broken-symmetry phase. However, second-order processes in $w/U$ and $w/E$ allow the system to reach states with more than two particles on a single lattice site (see \cite{PhysRevB.66.075128}). 
Our simulations show that the probability for having three particles on a single lattice site at times $t=4 \hbar / z$ of the order $10^{-6}$. It follows that this effect can be safely neglected  for an on-site interaction $U \gtrsim 8 \Delta$ on experimentally relevant timescales. With typical lifetimes of atoms in an optical lattice being on the order of seconds, an effective hopping rate $z/\hbar \sim 1$ Hz should be sufficient for observing the saturation of the correlation function $C(t)$. Further assuming a microscopic hopping rate of $w=15$ Hz, suppression of high-frequency oscillations yields a single-particle excitation energy $\Delta/\hbar = 450$ Hz. Finally, an on-site interaction $U/\hbar = 3.6$ kHz results in a stable broken-symmetry phase for the duration of the experiment. 

An important aspect is that throughout this paper calculations used an initial state of delocalized excitations with finite center of mass momentum $Q$ in order to maximize the correlation function $C(t)$. 
This behavior may seem counterintuitive, as one would expect a localized excitation to yield stronger correlations. However, localization in configuration space gives rise to a flat momentum distribution; with $C(t\gg \hbar/z) \rightarrow 0$ for $|Q| \le Q_{\rs c}$, this averaging over center of mass momenta then results in a decrease of the correlation-function by a factor $\sim 0.17$.
Still, the finite-$Q$ initial state can be composed of localized excitations via superposition, i.e 
\begin{align}
|\psi_Q \rangle = \sum_{j \in 2 \mathbb{N}}  e^{i \frac{Q}{2} j}  a_{j-1}^\dagger a_j   | \text{BSP} \rangle 
\end{align}
with $ | \text{BSP} \rangle $ denoting the broken-symmetry phase with two particles on every even lattice site. The factor of two in the phase then accounts for the two-site hopping of fractional excitations in the microscopic model.

We thank J. Simon and J. von Delft for fruitful discussions. Support from the DFG (Deutsche Forschungsgemeinschaft) within SFB/TRR21 and FOR 801 is acknowledged. Ian McCulloch acknowledges support from the Australian Research Council Centre of Excellence for Engineered Quantum Systems and the Discovery Projects funding scheme (project number DP1092513).
This research was supported in part by the National Science Foundation under Grant No. NSF PHY11-25915.

\end{document}